Article

# Blockchain-Based Distributed Patient-Centric Image Management System

**Mohamed Yaseen Jabarulla and Heung-No Lee \***

School of Electrical Engineering and Computer Science, Gwangju Institute of Science and Technology, Gwangju 61005, Korea; yaseen@gm.gist.ac.kr

\* Correspondence: heungno@gist.ac.kr

**Abstract:** In recent years, many researchers have focused on developing a feasible solution for storing and exchanging medical images in the field of health care. Current practices are deployed on cloud-based centralized data centers, which increase maintenance costs, require massive storage space, and raise privacy concerns about sharing information over a network. Therefore, it is important to design a framework to enable sharing and storing of big medical data efficiently within a trustless environment. In the present paper, we propose a novel proof-of-concept design for a distributed patient-centric image management (PCIM) system that is aimed to ensure safety and control of patient private data without using a centralized infrastructure. In this system, we employed an emerging Ethereum blockchain and a distributed file system technology called Inter-Planetary File System (IPFS). Then, we implemented an Ethereum smart contract called the patient-centric access control protocol to enable a distributed and trustworthy access control policy. IPFS provides the means for decentralized storage of medical images with global accessibility. We describe how the PCIM system architecture facilitates the distributed and secured patient-centric data access across multiple entities such as hospitals, patients, and image requestors. Finally, we deployed a smart contract prototype on an Ethereum testnet blockchain and evaluated the proposed framework within the Windows environment. The evaluation results demonstrated that the proposed scheme is efficient and feasible.

## 1. Introduction

Transition to electronic management of health records has necessitated practitioners and their patients to make use of several new acronyms such as electronic medical records (EMRs), electronic health records (EHRs), and personal health records (PHRs) [1]. These health records usually contain medical images and patient information, such as physician name, personal statistics (e.g., age and weight), home monitoring device data, and other data processed by practitioners in a text format. Medical images and patient information are stored and maintained by different hospitals, even when being related to the same patient. Current technologies for transferring medical images and patient information are deployed on centralized data centers that are deemed inappropriate due to privacy, accessibility, storage, and security concerns. Over recent decades, medical record data breaches within large medical data centers create additional difficulties for all companies seeking to develop medical image processing applications [2].

Recently, the blockchain technology, e.g., Bitcoin [3] and Ethereum [4], has become one of the most important research topics, not only in the finance industry but broadly across the field of information technologies due to its decentralized nature. Healthcare-based blockchain applications have been gaining particular attention in terms of applying them to enable interoperable sharing the real-time data among providers, payers, and patients [5,6].

Public blockchain technology is an open distributed ledger that stores all transaction details in blocks [3]. A typical blockchain consists of a directed acyclic graph (DAG) structure, where each block is linked with the previous block by a hash. Information stored in each block is public and cannot be easily deleted nor modified. Therefore, a blockchain is considered to be a decentralized method to facilitate verifiable exchanges of transactions between any two entities efficiently and permanently. Timely verification and recording of transactions are possible without the necessity in a centralized intermediary. A blockchain has such advantages as being tamper-proof and capable of protecting information against integrity-based attacks.

A significant problem with regard to storing medical images and records in a blockchain is the size of the content. For example, as of November 2020, the size of the Bitcoin blockchain reached 362.51 GB [7]. This is the result of data accumulation over the past ten years at a growth rate of 1 MB every 10 minutes since Bitcoin was launched in 2009. There are approximately 1000 transactions in a block. Thereby, a single transaction has the order of 1 KB. The size of medical images corresponds to the orders of magnitude larger than those a public blockchain can offer [8]. To solve the problem of decentralized storage, the Protocol Labs [9] created a distributed web called Inter Planetary File System (IPFS). IPFS was designed to enable a content-addressable, peer-to-peer (P2P) technology to share and store hypermedia in a distributed file system. Several other decentralized storage systems were developed, such as storj, swarm, and sia [10]. IPFS has an advantage of being compatible with other blockchain networks by offering an off-chain storage solution. IPFS provides permanent, smarter, and faster web services to distributed data access systems.

However, several obstacles exist in terms of storing sensitive medical images over these distributed storage solutions, such as unauthorized access and privacy concerns with regard to patient images. Namely, the ability to manage big data across general practitioners, hospitals, patients, and medical institutes without significant exposure to the risk of privacy breaches is essential. Another important aspect of a confidential and secure storage system is the ability to reduce the cost and restrictions of medical image acquisition by eliminating the need in centralized parties [11]. Therefore, the following research question is formulated:

"How can we design a patient-centric distributed architecture for the purpose of medical image storage and sharing, while simultaneously addressing the concerns about privacy, security, access flexibility, and costs?"

To answer this question, we propose a proof-of-concept (POC) design for a distributed framework called a patient-centric image management (PCIM) system that is a blockchain-based architecture designed to facilitate secured patient-centric access and storage of encrypted medical images within an open distributed network.

The contributions of this paper are as follows:
1. We provide a brief overview on the structure of the proposed PCIM system and illustrate interactions among different components of the system.
2. We propose a patient-centric access control protocol using a smart contract (PCAC-SC). Specific functions are considered to transmit information in and out of the Ethereum blockchain and give access privileges between entities.
3. We implement a framework to test feasibility of the concept. To this end, we have developed a PCAC-SC prototype on an Ethereum test network. We have published the related source codes online.
4. We verify the functionality using test cases and analyzed the capabilities of the proposed framework based on the following performance parameters: image access time, cost for executing functions, time taken to record PCAC-SC events in the blockchain, average block size, and average gas consumption.

The rest of the paper is organized as follows. In Section 2, we discuss the state of medical image sharing. The system components of the proposed framework are described in Section 3. An overview of the proposed PCIM system and PCAC-SC is presented in Section 4. Implementation, verification and analysis of the proposed system are described in Section 5. Section 6 discusses the advantages, limitations, and future research directions. Finally, Section 7 concludes the work.

## 2. Related Work

The practice of medical health record registering and sharing has changed considerably in the past 20 years, largely because of strict practice standards, the use of complex technologies, and accurate diagnosis and treatment. Medical images are typically shared on CDs or DVDs shipped between hospitals, physicians, and patients to conclude on diagnosis, however, applying this technology might lead to damage or interception of medical images resulting from patient or physician errors [12]. To overcome the shortcomings of physical media transfer, an internet-based standard communication technology called digital imaging and communications in medicine (DICOM) [13] was introduced to share, and store medical images across various healthcare enterprises. The two main components of DICOM standard are a DICOM file format and a network communication protocol which uses TCP/IP to communicate between systems. A DICOM file format consists of header tags and image data sets embedded into a single file which is unqualifiedly editable. Thus, the DICOM standard does not provide transmission security nor data protection [14]. The electronic transmission of DICOM medical images was developed by the Radiological Society of North America (RSNA) based on the image-sharing network (ISN) [15]. However, the ISN architecture employs picture archiving and communication systems (PACS) based centralized image storage, where images from multiple imaging modalities [16] are indexed by a cryptographic hash and managed by a third-party clearing house. The researchers [17] found that default accounts, cross-site scripting, and vulnerabilities in the web server could lead to breaches in PACS access and permanent modifications of medical images. The existing infrastructure design raises concerns regarding the use of third-parties and a centralized network.

Recently, several researchers focused on developing a framework that combine a cloud service and a blockchain for the purpose of medical health record sharing. The authors in [18] presented a specialized blockchain-based system for dermatology. Patients can access encrypted images and selectively share medical records using a private digital key. The authors discussed the possibility of allowing machine learning algorithms to access various images stored on the blockchain network to drive the optimization of computer-assisted analysis, but the scalability and cost effectiveness issue must be considered before standardizing this technique. In [19], the authors designed a breadcrumb mechanism for a medical record search known as MedBlock. Breadcrumbs were aimed to record addresses of blocks containing the patient-related data. Unfortunately, these solutions are not applicable to the process of searching the data over the blockchain due to an increase in the fragmented data. The authors in [20] proposed MeDShare, a hybrid cloud-based sharing solution for EHRs that is based on a centralized cloud server provider. Then, this external server was replaced by two decentralized networks called MedChain [21]. In the concept of MedChain, the authors proposed a session-based data sharing scheme and a digest chain structure implemented using an immutable blockchain and the mutable P2P storage architecture. However, the possibilities of tampering and manipulating stored patient health records are at high risk due to the mutable P2P storage architecture. In [22], a blockchain-based cross-domain image-sharing framework was proposed. However, no attempt to address privacy concerns has taken to facilitate sharing images through a blockchain.

## 3. System Components

In this section, we present the description of main components represented in the proposed PCIM system.

*3.1. Ethereum Blockchain*

Ethereum [7, 28] was developed based on the Bitcoin system and incorporated a programmable smart contract (SC) platform. In other words, SC is a computer program that stores rules for negotiating the terms of a contract. Programs can autonomously verify and execute contract-related agreements, thereby, reducing the cost of constructing and managing a centralized database. SC employs the Ethereum virtual machine that allows users to run SC within the blockchain network. In general, the fee mechanism of the Ethereum system depends on the value of gas [4]. A certain amount of gas is required to execute a SC and perform a transaction. A digital currency can be used to purchase gas. The actual transaction cost is defined as follows: Ether = gas used × gas price.

The Ethereum platform consists of two types of accounts: external owned accounts (EOAs) controlled by private keys and contract ones controlled by the contract code. EOAs are used to execute a transaction sending ether or to trigger execution of SC. An Ethereum transaction includes parameters such as recipient address, gas price, gas limit, ether values, account nonce, sender signature, and endpoint of the medical image. The Ethereum blockchain has an associated state database based on a Merkle-Patricia tree structure similar to IPFS objects. Therefore, we can model a blockchain using IPFS for more secure off-chain and on-chain storage of medical images. In the proposed scheme, we implemented the PCAC-SC protocol using an Ethereum blockchain to enable transparent controlled access, so that malicious entities could not access the medical images without patient authorization.

*3.2. IPFS Storage*

IPFS is a content-based peer-to-peer (P2P) protocol in which each medical image file is assigned with a unique fingerprint denoted as a cryptographic hash. Addressing the hash is applied to make the contents immutable [9]. The IPFS file storage structure consists of a Merkle DAG that combines Merkle trees with a DAG. The key feature of IPFS in terms of the proposed system is to access medical images through the content addressing approach, rather than location-based addressing one. Therefore, IPFS allows reducing the bandwidth cost, increasing the image download speed, and distributing a large volume of data with no duplication, in which allows achieving storage savings. The data structure for storing a file is an IPFS object, which consists of data and links. A single IPFS object can store up to 256 Kb of the unstructured binary data. If a file is larger than 256 Kb, it is split into and stored as multiple IPFS objects with an empty object containing links to all other objects of the image. Therefore, IPFS is an immutable storage mechanism; modifying a file will change the hash value. To update a file, IPFS uses a version control system called Git [24], which creates a commit object, when a file is added to the IPFS network; this approach allows tracking all file versions. When an update is made to a file, a new commit object is created as a link to a new object to interconnect with an older commit object version of that file.

*3.3. Securing Medical Images*

We encrypt the sensitive medical images before uploading to the global IPFS network in order to prevent unauthorized access. The participants can view the sensitive medical images securely by swapping encryption keys. This ensures data originality, ensures data security, and prevents data from being leaked to irrelevant users and being subject to malicious attacks such as eavesdropping, phishing, and brute force attacks [25].

The medical image is encrypted using the OpenPGP (Pretty Good Privacy) protocol [26]. OpenPGP is a specific implementation of asymmetric encryption that is used to define standard formats for encrypted messages, signatures, and certificates with the purpose of exchanging public keys. Therefore, a pair of asymmetric keys, a public and private one, is generated. The public key is shared openly without compromising the security, while the private key must be kept private. It is owned by the patient secretly and is used to decrypt the image. The advantage of applying this encryption technique is that using the private key, a digital signature of an image is created to verify its authenticity in the event of a malicious attack.

**4. Overview of the PCIM System**

In the proposed PCIM system, medical images are not stored in the blockchain to avoid scaling to the unmanageable size and thereby, a resulting blockchain bloat. Therefore, in the present study, we utilized the Ethereum blockchain for the proposed POC framework to efficiently manage the identity database and access control across participants. This action allows reducing the fees associated with storing images and managing the related database state. The fundamental purpose of this system is to provide distributed immutable on-chain and off-chain storage to facilitate patient-centric management for complex health record data. Figure 1 illustrates the blockchain ledger data structure with a PCIM data field added, as it is designed to store the data that patients want to include in a transaction.

| Account nonce |
| Gas price |
| Gas limit |
| Ether |
| Recipient |
| PCIM data |

**Figure 1.** Blockchain ledger data structure.

In the proposed scheme, the PCIM data field contents include such information as an image hash value (endpoint of an encrypted medical image), patient addresses, timestamp, encryption public key, image description, and a block hash to form an unchangeable record, as each block is linked with the hash of its previous blocks to connect and verify transactions. Every block is updated in the ledger after transactions are approved and recorded by a patient in the network. A transaction consists of a part corresponding to the ledger content signed and sent by a patient to execute SC by paying ether. Then, transaction validation is performed by the selected and approved consortium. As the blockchain is implemented in the healthcare ecosystem, participants seek to achieve decentralizing the process of medical data management. The overall architecture of the PCIM system framework is illustrated in Figure 2. As it can be seen, it consists of Ethereum and IPFS networks. The Ethereum network is comprised of PCAC-SC and of a blockchain ledger to manage identity and access control within the network. The resulting encrypted medical images are stored in the IPFS network. We discuss the participant interactions with system module in the following subsections (See Figure.3).

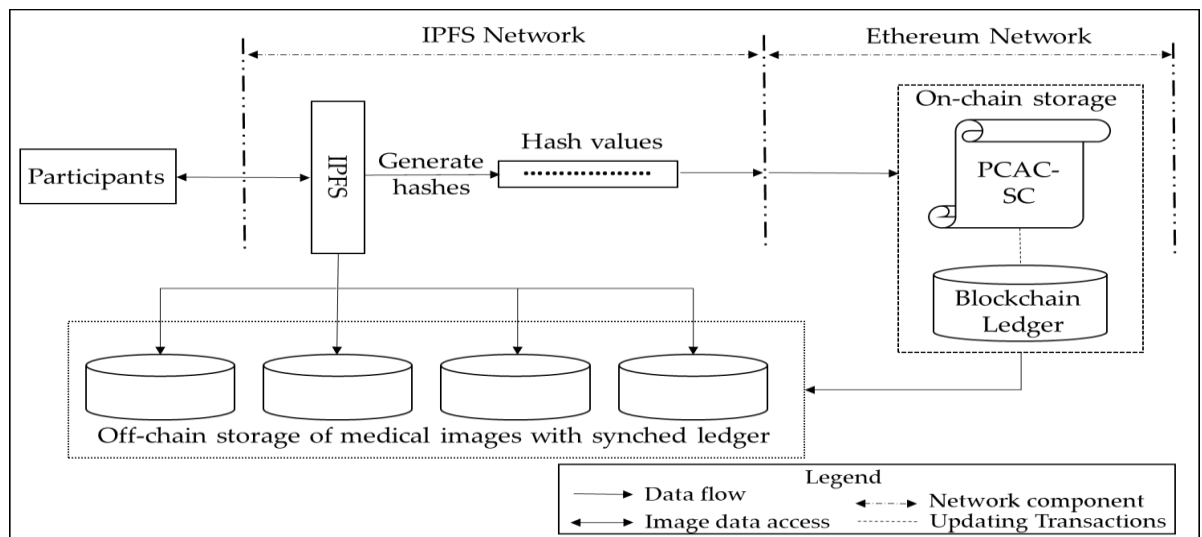

**Figure 2.** Architecture of patient-centric image management (PCIM) system. The architecture component split into two main decentralized modules: IPFS and Ethereum network.

*4.1. System Model*

The participants of the proposed PCIM system are defined below:
**Patient:** Patients are the owners of their medical images. A patient is required to create PCAC-SC and store this SC in the Ethereum blockchain. The patient is responsible for defining the access right to the images in the IPFS network. This definition is done within his/her own PCAC-SC.

**Radiologist:** A radiologist is able to generate medical images for a patient. The prime responsibility of the radiologist is to upload the patient encrypted medical images to the IPFS network and to verify the patient initial transaction on blockchain.

**Image Requestors (*IRs*):** Doctors, medical institutes, research groups, insurance companies, and general practitioners interested in accessing patient medical images are all considered as image requestors *IRs*. The patient can grant access privileges to any *IRs* based on the authorization policy defined in PCAC-SC.

*4.2. Ethereum Network: PCAC-SC Protocol*

We use the Ethereum SC to enforce access control policy on patient medical image on-chain contents. The SC stored in the Ethereum blockchain designed to contain unique image id, permissions, metadata and image integrity of the individual patient. The PCAC-SC protocol uses special functions to provide information about the blockchain and image access privileges for *IRs*. Furthermore, protocol helps in tracking all the on-chain activities of all participants. The functions of SC are triggered by a patient and *IRs* entity using their own Ethereum addresses. Thus, transaction costs reduced by using embedded protocols to reduce administrative burdens and remove intermediaries.

All triggered functions are stored within the blockchain ledger as events to allow the entity keeping track of the transaction details. This enables transparency in the triggered functions and maintains the anonymity of patients by displaying only events stored in the blockchain. In this framework, we used a single variable, and the following functions:

*msg.sender*: the address variable of the owner who interacts with SC.

*create_contract():* this function is created and executed only by a patient to issue corresponding roles for *IRs* and related information for accessing medical images. As shown in Algorithm 1, this function takes as input a patient's encrypted medical image hash value $h(\overline{I_p})$, blockchain address $\Phi_P$, image description $\Delta_P$ and the timestamp when the function was executed by SC.

---
**Algorithm 1:** *create_contract()*

**Input:** $h(\overline{I_p})$, $\Phi_P$, $\Delta_P$

**Output:** bool

1: **if** *msg.sender* is not $\Phi_P$ **then**
2: throw;
3: **end**
4: mapping $h(\overline{I_p})$ to ($\Phi_P$) and add it to ledger
5: **return** true;

---

*requesting_access():* this function is executed by *IRs* to obtain access permission from the patient. *IRs* includes as input the patient blockchain address $\Phi_P$ and *IRs* public key $K_{IR}^+$ to encrypt medical images and additional information, such as usage notes as shown in Algorithm 2.

---
**Algorithm 2:** *requesting_access()*

**Input:** $\Phi_P$, $K_{IR}^+$, Notes

**Output:** bool

1: **if** *msg.sender* is not $\Phi_{IR}$ **then**
2: throw;
3: **end**
4: call PCAC-SC ();
5: **if** new_IRs_address $\Leftarrow$ approved **then**
6: **return** true;

```
7: else
8: if new_IRs_address ⇐ not approved then
9: return false;
10: end
```

*approve_IRs()*: this function can only be executed by the patient. As shown in Algorithm 3, it grants/denies access permission by using as input the *IRs* blockchain address $\Phi_{IR}$, *IRs* public key $K_{IR}^+$, and notes from *IRs*. The input notes contain relevant information such as the expiration date and message for requestors.

**Algorithm 3:** *approve_IRs()*

**Input:** $\Phi_{IR}$, Notes
**Output:** bool
```
1: if msg.sender is not Φ_P then
2: throw;
3: end
4: if Φ_IR exist then
5: return false;
6: else
7: authorize_User[ Φ_IR ] ⇐ true;
8: mapping h(Ī_p) to ( Φ_IR ), and add it to ledger
9: return true;
10: end
```

*trace_authorization()*: this function executed by *IRs* and patients to track the history of approved or disapproved requestors in the blockchain. Thus, participant authenticity to access patient medical images verified by calling this function. The following algorithm used by the *trace_authorization()* function.

**Algorithm 4:** *trace_authorization()*

**Input:** $\Phi_P$, $\Phi_{IR}$
**Output:** bool
```
1: if msg.sender is not Φ_IR then
2: throw;
3: end
4: if Φ_IR exist then
5: return true;
6: else
7: return false;
8: end
```

*remove_IRs()*: this function takes the approved *IRs* blockchain address $\Phi_{IR}$ as input and removes *IRs* from SC upon successful execution of a function by the patient as defined in Algorithm 5. Consequently, SC is updated. Therefore, the removed *IRs* has no privilege to access the medical image contents stored on-chain. Note that, this function used to record the log of removed entity in the blockchain as a proof. Thus, the participants are legally not allowed to access the medical images.

**Algorithm 5:** *remove_IRs*

```
Input:  Φ_{IR}
Output: bool
1: if msg.sender is not  Φ_P  then
2: throw;
3: end
4: if  Φ_{IR}  is not exist then
5: return false;
6: else
7: authorize_User[ Φ_{IR} ] ⇐ false;
8: return true;
9: end
```

*4.3. IPFS Network*

IPFS is used to store encrypted medical images that contain the encrypted patient information in an open distributed storage system, in which images can be exchanged using a hash string path. The paths work similarly to the traditional uniform resource locator used in the web. Therefore, all patient images are always accessible through their hash.

The radiologist uploads medical images of the patient to the system and uses a patient public key to encrypt the images: thereafter, only the patient can decrypt them. Medical image contents are signed by Ethereum private keys of the patient and then, are stored in the blockchain. Therefore, other entities can check the authenticity and integrity of the image ownership using the content hash and digital signature in the blockchain. In IPFS, files can be accessed even if the host node is offline, as they are located in multiple locations for redundancy. SC enforces access control only to the on-chain content stored on blockchain. It controls access to the medical image in the IPFS network in terms of image file attributes, that help in tracking all the on-chain activities of participants. Moreover, The PCIM system protects the off-chain patient data using the security and privacy feature. Therefore, combining IPFS and the blockchain allows building a permanently addressable on-chain and off-chain data storage that can be linked securely to other significant systems or databases in the world, thereby, forming a global healthcare network.

*4.4. System Interaction*

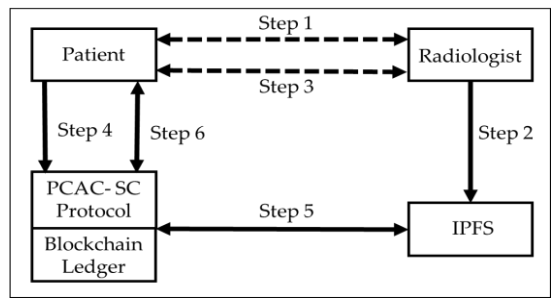

**Figure 3.** Interaction model of the PCIM system.

Figure. 3 illustrates the process of how a patient and a radiologist interact between each other in the part of the proposed PCIM system, where medical image storage and sharing are performed. First, the patient undergoes the medical image examination performed by the radiologist. A medical image $I_P$ of the patient is produced. The patient seeks to have it protected and to maintain the ownership of this image. Consequently, to address this issue, the radiologist encrypts the initial medical image and obtains encrypted image. Thereafter, the radiologist obtains the hash of the encrypted image $h(\bar{I}_P)$ from the IPFS network and provides the patient with $h(\bar{I}_P)$ for the reference purpose. $h(\bar{I}_P)$ is stored in the blockchain, while the encrypted medical image $\bar{I}_P$ is stored in IPFS.

Owing to the fact that the image was encrypted, the patient medical image $I_p$ is accessible only to those who have the decryption key and thereby, it is protected from unauthorized access.

As presented in Figure.3, the exact protocol for this interaction is explained in detail as follows:

- Step 1: *Offline interaction between the patient and the radiologist*
    Step 1.1: The patient requests the radiologist to store his/her medical image.
    Step 1.2: The radiologist asks the patient to provide its encryption key.
    Step 1.3: The patient generates a pair of encryption keys: public $K_P^+$ and private $K_P^-$.
    Step 1.4: The patient sends the public key $K_P^+$ to the radiologist through a secure communication medium for creating image authentication and encrypt the original medical image.
    Step 1.5: $K_P^-$ is protected and kept safe by the patient.
- Step 2: The radiologist encrypts with $K_P^+$ while concealing the patient private information on a medical image. Encrypted image $\bar{I}_P$ is uploaded to the IPFS network, which returns a hash $h(\bar{I}_P)$ to the radiologist.
- Step 3: The radiologist shares $h(\bar{I}_P)$ through a secure communication medium with the patient.
- Step 4: The patient creates a contract using the PCAC-SC protocol and executes it.
- Step 5: The created contract function signs a transaction on the Ethereum blockchain along with patient public key ($\Phi_P^+$), $h(\bar{I}_P)$, time, image description ($\Delta_P$) such as patient blockchain address ($\Phi_P$), and an imaging modality from which the data are obtained (e.g., CT, US, MRI, etc). This transaction is verified by the radiologist and included in the blockchain. This verification process prevents multiple entities from executing *create_contract()* function on the same image hash.
- Step 6: The patient owns the medical images within the PCIM system. The patient can access, audit, prove the ownership, and authorize any other *IRs* (e.g., doctors, medical institutes, research groups and general practitioners) to use their medical images based on PCAC-SC. We discuss the PCAC-SC interaction sequence in Sub-Section 5.2.

In summary, a blockchain transaction consists of the following contents signed by a patient to represent the ownership of the transaction contents:

$$\{\Phi_P, h(\bar{I}_P), \Delta_P\}_{\Phi_P^{-1}}$$

where the part given inside the parenthesis, { }, is the content signed under the Ethereum blockchain private key $\Phi_P^{-1}$ of the patient.

*4.5. Medical Image Sharing*

Medical image sharing between a patient and an image requestor is based on PCAC-SC protocol. For example, consider a new image requestor interested in accessing the patient medical images for research purposes.

The protocol for gaining an access based on PCIM system is as follows:

- Step 1: Requestor shares $K_{IR}^+$ a public key using *requesting_access()* a SC function.
- Step 2: Patient downloads the encrypted image from the IPFS network using the IPFS hash value.
- Step 3: Patient decrypts the encrypted image with patient's own private key $K_P^-$.
- Step 4: Patient obtains the requestor's public key by providing the requestor's blockchain address.
- Step 5: Patient encrypts the original image with the requestor's public key $K_{IR}^+$ and uploads the encrypted image to the IPFS network.
- Step 6: Patient signs a transaction on the blockchain along with the requestor's public key, the patient's public key and the IPFS hash value using *approve_IRs()* function.
- Step 7: The image requestor is able to retrieve the medical image using the IPFS hash value and decrypts with his/her own private key $K_{IR}^-$. In this way, medical images are shared between the patient and the requester.

In this protocol, if the same medical image has been requested by multiple requesters, then patients need to encrypt the image with requester's public key, and upload the same image to the IPFS network. This process may result in a waste of storage resources. To overcome this issue, the participant of the network follows a strict regulation for limiting the number of approvals, such that only a certain number of image requests are approved by the patient within a particular period of time. Placing such limitation can help reduce the waste of storage space in the IPFS network.

## 5. Evaluation

### 5.1. Experimental Setup

A POC design of the PCIM system was developed to test and evaluate its performance. The experiment was conducted using a Windows 10 desktop with an Intel® Core ™ i5-6600 processor at 3.30 GHz. PCAC-SC was implemented in the remix IDE [27] using Solidity [28] programming language. We deployed the program within the private Rinkeby test network using MetaMask [29]. This test network allows us to verify and optimize the prototype before implementing in a public blockchain. We initialized IPFS using go-ipfs [30] and uploaded an encrypted medical image to the IPFS network from a local computer. This operation returned a unique hash value linked to the uploaded medical image. Thereafter, we updated transactions on the blockchain using *create_contract()* function by defining the IPFS hash, patient Ethereum public key, and the basic medical image description. Once the block was approved, transactions were stored in the blockchain.

The complete prototype code of PCAC-SC is published in our GitHub repository [31]. The contract deployed on the test network has the following address:

$$0x5575805E19b4807974Be0B77Fd9d385D4A0e6d1E$$

Transactions on each function can be seen using the above address at the Rinkeby Etherscan website [32].

Figure 4 illustrates such parameters as the block/timeline, functions, and event sequence defined in the PCAC-SC protocol for granting and revoking permissions between a patient and image requestor *IRs* entities. To allow for better understanding of this access sharing sequence, we consider an example of two *IRs*: a doctor ($IR_1$) and a general practitioner ($IR_2$) who is interested in accessing a patient medical image.

The patient executes *create_contract()* function by signing the blockchain contents (see Sub-Section 4.4). This function allows $IR_1$ and $IR_2$ to participate by calling the request function in the PCAC-SC protocol defined by the patient. Each of the entities has its own Ethereum address to perform the operations.

In Figure 4, blocks from 2 to 7 illustrate the access privilege scenario. $IR_1$ and $IR_2$ send a request to access medical images using *request_access()* function that is represented by block 2 and block 3. Block 4 and block 5 show that the patient is able to grant and deny the access by using the *approve_IRs()* function. In Figure. 4, block 6 depicts an event that $IR_1$ authorized to access image, since in the block 4 $IR_1$ image request accepted by the patient. Thus, block 6 represents the message sequence of *trace_authorization()* function, which is used to trace the history of approved and disapproved events $IR_2$ of the image requestors. In Figure.4, block 7 illustrates revoking the permission of $IR_1$ by calling *remove_IRs()* function, which can be executed only by the patient. The details on execution of each function are stored in the blockchain as an event to help the participants to keep track of their transaction details.

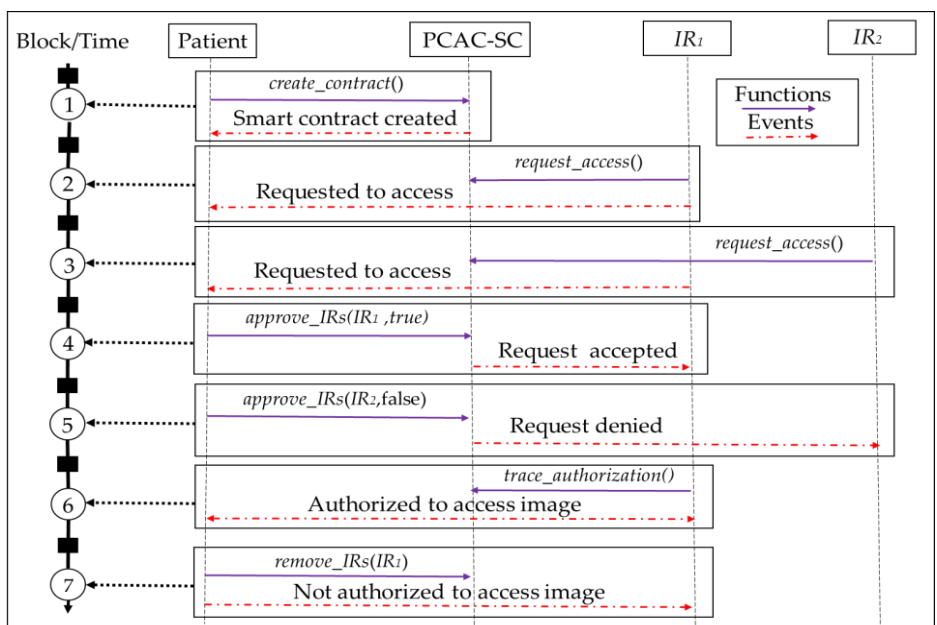

**Figure 4.** Access sharing sequence. The blockchain/timeline is shown on the left, pointing with dotted arrows for reference. The purple, and red arrows represent interactions between entities.

*5.2. PCAC-SC Verification*

We verify the access sequence and interaction between entities by testing two main functions for brevity. We choose *approve_IRs()* for the accept/deny permission to access a medical image and *trace_authorization()* to verify access privileges for a given Ethereum address. Figure 5 shows that the approved *IRs* and trace authorization functions provide the following test cases: request accepted, request denied, authorization success, and authorization failed. The results obtained by testing the case 1 to case 4 is shown in Figure 6-9. The Figures 6-9 depict the summary of the transaction event log stored in the blockchain after the successful execution of the functions.

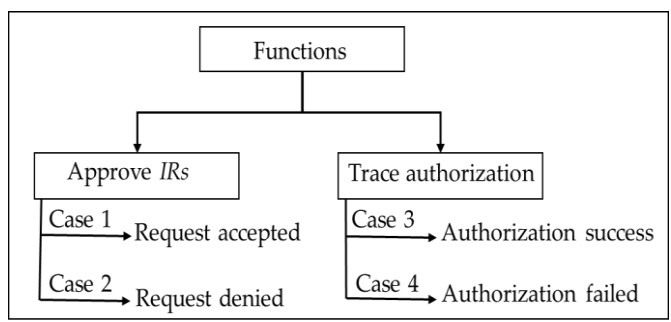

**Figure 5.** PCAC-SC validating functions and testing cases.

To test the prototype, we consider the following Ethereum address and IPFS hash of the medical image:

- Patient Ethereum address: *0x5575805E19b4807974Be0B77Fd9d385D4A0e6d1E*
- $IR_1$ Ethereum address: *0xdD870fA1b7C4700F2BD7f44238821C26f7392148*
- $IR_2$ Ethereum address: *0x583031D1113aD414F02576BD6afaBfb302140225*
- IPFS hash: *QmNaS5gQzoPxr3S2n6T6BsFuVRmMFwpohLVFfAFrU8gyTq*

5.2.1. Testing an approved IRs function

In this testing, we consider the first case, where a patient approves the $IR_1$ address to access medical images by mapping with the IPFS hash value. Events *requestaccepted* and *approved* were triggered by *approved_IRs()* function, and $IR_1$ gained access privileges to a patient medical image. The event is stored in the blockchain as shown in Figure. 6.

> {"event": "Requestaccepted",
> "patient":"0x5575805E19b4807974Be0B77Fd9d385D4A0e6d1E",
> "info": "approved by patient."}
> {"event": "Approved",
> "requester":"0xdD870fA1b7C4700F2BD7f44238821C26f7392148",
> "info": "Authorized to access image"}

**Figure 6.** Case 1: event log for approving $IR_1$ address to access a patient medical image.

Figure 7 shows the second test case, where a patient denies $IR_2$ request to access medical images. This function triggers two events *requestdenied* and *reason* for rejecting by the patient.

> {"event": "Requestdenied",
> "patient":"0x5575805E19b4807974Be0B77Fd9d385D4A0e6d1E",
> "info": "Failed to be approved by patient"}
> {"event": "Reason",
> "requester":"0x583031D1113aD414F02576BD6afaBfb302140225",
> "info":"Need more detailed information to access my image"}

**Figure 7.** Case 2: event log stored in the blockchain for denying access to $IR_2$ address.

5.2.2. Testing trace authorization function

Here, we test the *trace_authorization()* function. This function is used to prove the ownership and trace history of the approved *IR's* in the blockchain. To verify authorization, let us consider that the $IR_1$ address is already approved. Patient and $IR_1$ Ethereum address are given as input to execute *trace_authorization()* function, and this triggers *authorizationSuccess* event. Figure. 8 shows the event log of the third test case where $IR_1$ address is authorized to access an image by the patient.

> {"event": "AuthorizationSuccess",
> "requester":"0xdD870fA1b7C4700F2BD7f44238821C26f7392148",
> "info": "Authorized to access image",
> "patient":"0x5575805E19b4807974Be0B77Fd9d385D4A0e6d1E"}

**Figure 8.** Case 3: event log stored in the blockchain. Information shows that the $IR_1$ address was authorized to access a patient medical image.

Figure 9 shows the log of *authorizationfailed* event invoked from SC. This event occurs due to the fact that the $IR_1$ address has been removed or has not been approved by the patient.

> {"event": "AuthorizationFailed",
> "requester":"0x583031D1113aD414F02576BD6afaBfb302140225",
> "info": "Liver image is not authorized to access",
> "patient":"0x5575805E19b4807974Be0B77Fd9d385D4A0e6d1E"}

**Figure 9.** Case 4: event log where the $IR_2$ address is not authorized to access a patient medical image.

*5.3. PCIM System Analysis*

In the previous sections, we have demonstrated how a medical image can be stored and shared in a decentralized network using the PCIM system. In this section, we analyze our proposed scheme on the basis of following performance parameters which are image access time, cost for executing functions, time taken to record PCAC-SC event in blockchain, block size, and average gas consumption.

5.3.1. Evaluating the Image Access Time

For this experiment, we set up the conventional cloud-based PACS (C-PACS) using PostDICOM [34], which uses Amazon S3 to store the medical images. We evaluated the efficiency of image access time by comparing the PCIM storage system with the C-PACS based on two parameters: the number of submitted images, and the size of the stored medical images. The measurement of the performance was based on the following metrics: upload and download time of medical images. We obtained anonymized sample images from the DICOM library [33] to verify the image access time in Seconds. The experiment was performed on our local computer with an internet download speed of 272.19 Mbps and an upload speed of 55.92 Mbps. The medical image size range from 1 Mb to 100 Mb is uploaded to our distributed IPFS network and conventional cloud network and then downloaded the images to the local computer. The upload time for PCIM is observed when medical images get stored into the IPFS network and download time is observed during the medical images accessed using the IPFS hash value stored in the blockchain. Figure 10 shows the time taken to upload and download the medical images using the PCIM and C-PACS storage. The line graph shows that the proposed system takes less than 1 Second to upload 4 to 115 medical images and takes 5.31 seconds to upload 100 Mb of medical images. Whereas, the C-PACS system takes nearly 8 Seconds for uploading 10 Mb of images and it continues to increase as the image size reaches 100 Mb. The bar graph in Figure 10 shows that within 6.19 Seconds 1142 medical images are downloaded by the proposed system and C-PACS takes 19.43 Seconds to download similar number of images. As evident from the figures, we can see that the medical image access time of the PCIM storage network outperforms the conventional cloud system. As expected, the proposed approach is faster due to its distributed operations. The traditional storage system is a complex process due to the centralized server, queued transactions, and privacy issues. In general, the experiments show that the IPFS based storage solution is robust and possible to access all the images faster and without any interruptions.

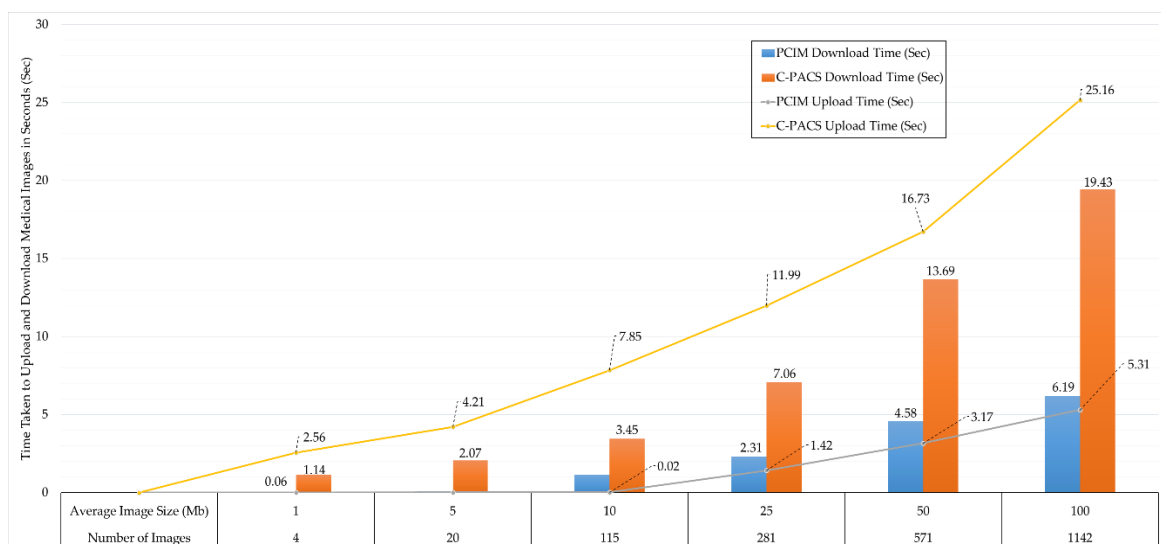

**Figure 10:** Showing the Average Time Taken to Upload and Download the Medical Images of Various Sizes by the Proposed and Conventional PACS Storage in Seconds (Sec). The Bar Graph Represents the Download Time and the Line Graph Represents the Upload Time of the Two Systems. The Dashed Lines Represent the Numerical Data Labels of the Line Graph.

5.3.2 Costs and Practicality

In our proposed system, we define the actual transaction cost to be Ether = gas used × gas price. Here, 'gas used' represents the constant computational cost. The gas price is adjusted by the network [4] to compensate for changes in the value of Ether. Thus, the total transaction cost (Ether) is kept relatively constant for the accessibility of health care data. As for the payer segment, every participant has to pay in gas for executing an operation in SC. Thus, the automated process of SC would cause significant cost savings for the patient.

In the implemented PCAC-SC prototype, we set a gas limit of 30,000, where each unit of gas is set equal to 2 Gwei. The total transaction fee in this scenario is 0.11 USD. Table I summarizes the cost of the executed operations in SC. The *create_contract()* function is implemented once with a cost of 0.025 USD. The *request_access()* function cost is 0.093 USD, which is higher than that of other functions due to the additional input bytes included during the function execution, such as those corresponding to the patient blockchain address and notes for the usage agreement. The overall costs can be decreased further if the size of the input data is minimal. However, these costs are still lower than those associated with buying a storage space from a third-party service or maintaining a database using a centralized system such as ISN [15], MedBlock [19], MeDShare [20], and MedChain [21].

**Table 1.** PCAC-SC Cost Analysis (gasprice = 2 Gwei, 1 ether = 187 USD).

| Function | Gas Used | Actual Cost(ether) | USD |
|---|---|---|---|
| *create_contract()* | 67394 | 0.000134788 | 0.025 |
| *requesting_access()* | 246908 | 0.000493816 | 0.093 |
| *approve_IRs()* | 170412 | 0.000340824 | 0.064 |
| *trace_authorization()* | 34266 | 0.000068532 | 0.013 |
| *remove_IRs()* | 59358 | 0.000118716 | 0.022 |

5.3.3. Transaction Efficiency

We performed the efficiency analysis of the PCAC-SC based on transaction storing time and gas consumption. The multi-bar graph in Figure 11 reports the deployment and execution time of PCAC-SC in Ethereum testnet using five different patients' Ethereum address. It is noted that storing *create_contract()* events takes more time than deploying the SC and storing the events of *requesting_access()*, *approve_IRs()*, *remove_IRs()*, and *trace_authorization()* functions. Our results show that most transactions were validated and written into a block within 4.98 to 16.97 Seconds. The access granting process took in significantly less time than the rest of the functions. Figure 12 shows the gas used for storing a number of transaction events in the blockchain and its block size. It has been noted that maximum gas (66.94%) was consumed by the fourth transaction since the block size is high (20629 bytes). The minimum gas (8.85%) was consumed by the seventeenth transaction, while the block size is low (3485 bytes). We observed that gas consumption increases and decreases based on the block size. Furthermore, the analysis shows that gas consumption decreases as the number of transactions increases. Gas consumption percentages in our experiment were varying but reasonably stable. Thus, the above analysis shows that the proposed scheme outperforms in terms of average time and gas consumption to store the events in the blockchain.

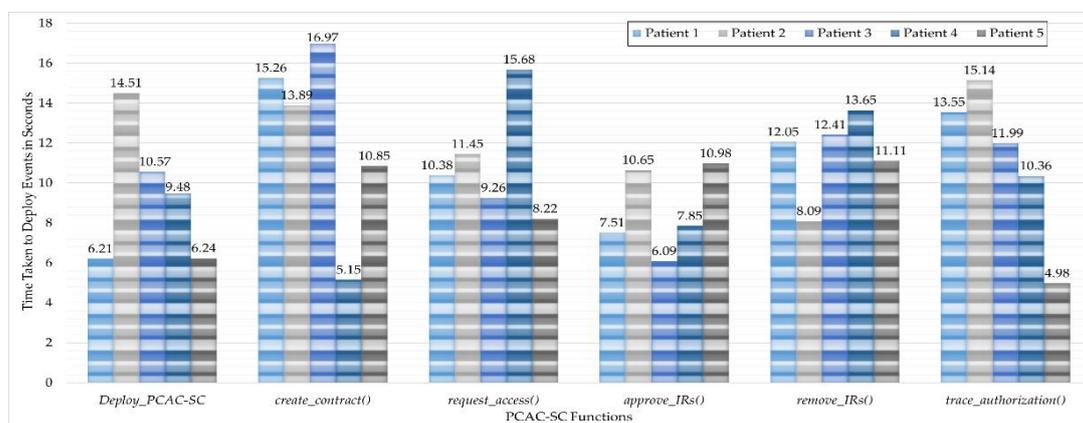

**Figure 11.** Average Time Taken to Record the PCAC-SC Events by using Five Patients' Ethereum Address.

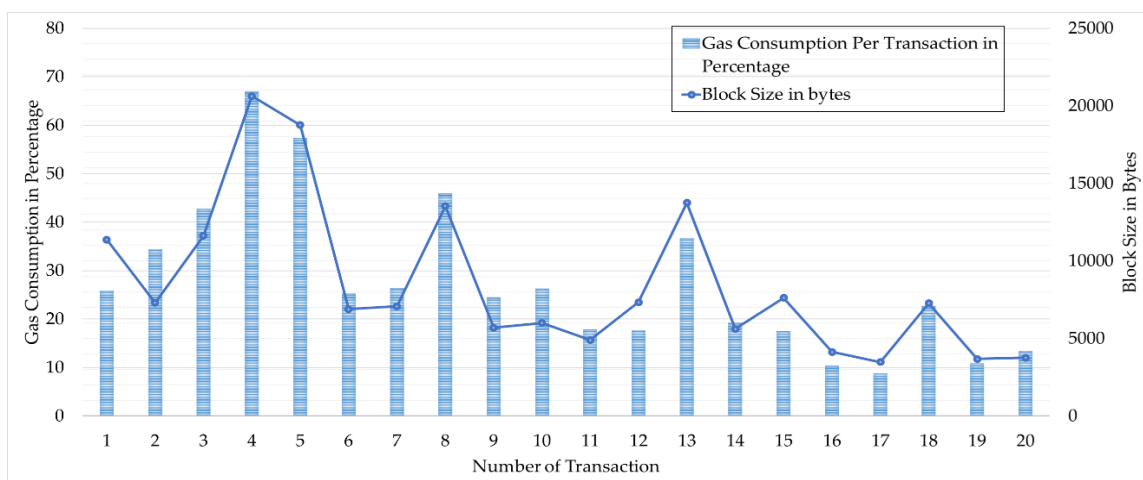

**Figure 12.** Gas Consumption and Block Size versus Number of Transaction of the PCAC-SC functions.

## 6. Discussion

The blockchain technology in PCIM system cryptographically protects the state of transactions of medical images and provides higher efficiency in terms of cost, event storage and image access overhead as discussed in the previous section. The use of IPFS allows constructing a high-throughput content-based storage model with content-addressed hyperlinks. In addition, the medical image migration time and retrieval time is faster compared to the conventional cloud based PACS. Furthermore, in Table 2 we provide a summarized comparison between the proposed framework using the ISN [15] and alternative blockchain-based medical health record management frameworks [24–26]. From this table, it can be seen that the proposed PCIM system has greater advantages comparing with the existing alternatives. Among them, studies [15,19,20] are based on centralized frameworks in which one central node failure causes a fail of the whole system. In contrast, in the framework proposed in this paper, every node is independent of each other, which ensures robust and efficient data access. The MedChain [21] uses a mutable P2P storage network, which has a high risk of data attacks and content duplication. The proposed PCIM system overcomes these disadvantages by using an IPFS-based storage in which medical images corresponding to the same content are not allowed being duplicated. This allows users to have full control of their medical images by ensuring guaranteed security, transparency, and data integrity. If the contents in a file stored within the IPFS network are not peered or active for a period of time, it is recycled by the garbage collector. Protocol labs understood this limitation of IPFS and build Filecoin as a complementary component, that turns cloud storage into an algorithmic market [35]. Furthermore, a participant must pin the image content to ensure that the content never gets deleted by the garbage collector. The image stays up indefinitely, as long as it is pinned on the IPFS network. Using the IPFS and blockchain system has the advantage of replacing current expensive storage systems (PACS) and centralized databases and of eliminating the recovery cost in the event of data breaches.

However, we note some limitations for a broader concern, due to the decentralized nature of our system, such as losing private keys. In some studies [36,37], the researchers introduced an efficient recovery mechanism using biometric data to create key pairs. This technique helps patient to securely store keys on their devices and recover the key in case they are lost. Furthermore, the medical images are not protected once the sensitive image is decrypted. It is difficult to identify the authorized recipients who attempt to tamper or manipulate the decrypted medical image. To overcome this issue, there are several data hiding techniques, i.e., watermarking, reported in the literature [38–40]. However, these data hiding techniques have not yet been clinically employed. In our future work, we will address the aforementioned limitations by considering the biometric signature [36] and steganography techniques [40]. Another limitation is that the encryption and the decryption of original medical images are performed manually on the IPFS network by the participants. Thus, in the future, the complexity underlying this action will be improved with a user friendly application interface for emergency access. Overall, the practical usefulness of the PCIM system depends on the participant experience. Decentralized data management is the groundbreaking of blockchain development, and it is more of a fascinating prototype of what health care technology could look like in the future.

Table 2. Comparison between the Existing and Proposed PCIM System

| Schemes | ISN [15] | MedBlock [19] | MeDShare [20] | MedChain [21] | PCIM system |
|---|---|---|---|---|---|
| Source Data Storage | PACS | Dedicated Servers | Cloud Server | Mutable P2P Storage | Immutable IPFS Storage |
| Source Data Encryption / Scheme | Yes / Not Mentioned | Yes / Symmetric Encryption | Yes / Not Mentioned | Yes / Asymmetric Encryption | Yes / Asymmetric Encryption |
| Type of Data | Medical Images | EMR | EMR | EHR | Medical Images |
| Server Attack Resistance | No | No | No | No | Yes |
| Tamper-Proof Database | No | Yes | Yes | Yes | Yes |
| Database Sharing Mechanism | PACS | Blockchain | Blockchain | Blockchain | Blockchain |
| Database Management | Centralized | Centralized | Centralized | Semi-Centralized | Decentralized |
| Smart Contract | No | No | Yes | No | Yes |
| Offline Data Access | Yes | No | No | No | Yes |

## 7. Conclusion

Patient medical images are the most valuable asset of any healthcare system's intelligence. Most of the time, these medical images are indeed scattered across different systems, and sharing them is influential for establishing effective and cohesive healthcare. In addition, a centralized hosting location of image data (e.g., cloud-based solution) can be a single point of a security attack. With growing recognition of the distributed nature of health services, attention has been increasingly focused on decentralized architectures and system interoperability. In this paper, we presented the POC design of the proposed PCIM system: an Ethereum blockchain and IPFS-based decentralized framework for storing and sharing medical images. Moreover, we introduced a new access management system called PCAC-SC that enables authorized entities to access the relevant blockchain data. The PCIM system facilitates a new way to improve the right of patients to perform self-determination regarding their medical images. We performed the experimental implementation to analyze and evaluate efficiency, rationality and feasibility of the proposed scheme. The proposed system facilitates patient access to an immutable medical database providing higher efficiency, data provenance, and effective audit while sharing medical images. The data storage and exchange model is also decentralized; therefore, necessity to involve third-party intermediaries and administrative structures is eliminated. Our future research goal is to deploy the proposed POC design in the public blockchain using real-time scenarios to form a global PCIM system, as well as to evaluate policies and regulations to adopt this emerging technology within healthcare enterprise.

**Author Contributions:** M.Y.J. and H.N.L. conceptualized and designed the research work; M.Y.J. performed the implementation and evaluation of the research work; M.Y.J. wrote the manuscript; H.N.L. performed investigation, made revisions to the manuscript and supervised the research work.

**Funding:** This work was partly supported by the "Practical Research and Development support program supervised by the GTI (GIST Technology Institute)" grant funded by GIST in 2019 (Development of Error Correction Codes PoW and Bitcoin/Ethereum Hardfork 1.0) and the National Research Foundation of Korea (NRF) grant by the Korean government (MSIP) [NRF-2018R1A2A1A19018665].

**Conflicts of Interest:** The authors declare no conflict of interest.